\documentclass{aa}  

\usepackage{graphicx}
\usepackage{parskip}
\usepackage{txfonts,textcomp}
\usepackage{multirow}
\usepackage{xcolor}

\usepackage[colorlinks=true,     linkcolor=blue, citecolor=blue, filecolor=blue, urlcolor=blue]{hyperref}

\begin{document} 


   \title{Laboratory experiments on the sublimation of methane through ice dust layers and applications to cometary activity}

   \subtitle{}
      \author{Carla Tamai
          \inst{1}
          \and
      Belén Maté
       \inst{2} 
        \and
       Stéphanie Cazaux
       \inst{1,3}
       \and
       Miguel Ángel Satorre
       \inst{4}
       }

   \institute{
               Faculty of Aerospace Engineering, Delft University of Technology, Delft, The Netherlands   
        \and Instituto de Estructura de la Materia, IEM-CSIC, Calle Serrano 121, 28006 Madrid, Spain \\
              \email{belen.mate@csic.es}
        \and
            Leiden Observatory, Leiden University, P.O. Box 9513, NL 2300 RA Leiden, The Netherlands\\
            \email{s.m.cazaux@tudelft.nl}
         \and
             Escuela Politécnica Superior de Alcoy, Universitat Polit\`{e}cnica de Val\`{e}ncia, 03801 Alicante, Spain \\
             \email{msatorre@fis.upv.es}
             }
   \date{Accepted: 12/05/2023 }
   
 
  \abstract
  {Comets are small celestial bodies made of ice, dust, and rock that orbit the Sun. Understanding their behavior as they warm up at perihelion unveils many pieces of information about the interior and general morphology of the ices hidden under the dust.}
  {The goal of this research is to study the sublimation of CH$_4$ through amorphous solid water (ASW), with a focus on the structural changes in water and the influence of a layer of indene (as a proxy of the crust) during a period of thermal processing, which we use in a controlled laboratory setting to simulate cometary environments.}
  {Ices at a  CH$_4$/H$_2$O abundance ratio of about 0.01 are deposited and layered, or co-deposited, at 30 K and are heated until 200 K (or 140 K) with a ramp of either 1 or 5 K/min. We use mass spectrometry and infrared spectroscopy to analyze the results.}
  {Depending on the heating ramp and type of deposition, the sublimation of methane through ASW varies, being lower in intensity and higher in temperature when the co-deposited structure is considered. When two temperature cycles are applied, the second one sees less intense CH$_4$ desorptions. When indene is placed above the ice mixtures, we find that the thicker its layer, the later the methane desorption. However, this later desorption sees a greater quantity of methane released due to water reorganization and higher desorbed material pressure.}
  {
  The structural changes of water ice drive volatile and hyper-volatile desorption because of the transition from high to low intrinsic density and transformation from amorphous to crystalline. This desorption indicates that such material has been deposited at low temperatures in agreement with previous theories on cometary ices formed in the pre-stellar cloud. During the two temperature cycles of our experiments, most of the released material is seen to be pristine and the processed part, if any, is of a negligible quantity, in agreement with dust--rock cometary studies.
  }

\titlerunning{Diffusion and sublimation of methane through ice dust layers to mimic cometary nucleus activity.}

   \maketitle
%
   
\section{Introduction}
Comets, and in particular their ices, retain key information on their formation conditions, 4.6 Gyr ago in the solar nebula. Water ice is a major ice constituent of comets, and its release is usually accompanied by the release of volatile compounds such as CO and CH$_4$, which are present in lower abundances  \citep{bockelee2004composition}. As comets approach the Sun, the release of water and volatiles increases until perihelion, when it then decreases as the comets move away from the Sun \citep{stern1999comet,hansen2016evolution,biver1997long,snodgrass2016perihelion}. The desorption patterns of water and volatile compounds as comets approach or move away from the Sun have been described by \cite{biver1997evolution,biver20021995} and \cite{ rubin2020origin} for comets Hale Bopp and 67P/Churyumov-Gerasimenko. These studies report desorption slopes that change while approaching or leaving perihelion. An explanation for the slope changes observed remains elusive.

To understand the processes involved during the desorption of cometary ices,  several experiments have been performed that have added to our knowledge of the morphology of cometary ices \citep{meech2004using} and of the diffusion and trapping of the volatile compounds contained within them. Some of these studies focus on the effect of porosity on the rate of outgassing of porous granular ice \citep{kossacki2021sublimation}, or the adsorption, desorption, trapping, and release of volatiles by amorphous solid water \citep{ayotte2001effect}. Other studies investigate the influence of carbon monoxide, carbon dioxide, methane, and ammonia on the sublimation of water \citep{kossacki2017sublimation}, or the migration of water through the internal ice layers of a comet to reach the surface \citep{pat2009experimental}. In some studies, the water ice is covered with dust layers in order to determine the effect of dust on diffusion or desorption; see for example \cite{gundlach2011outgassing}. Another study from  \cite{krause2011thermal} focused on the thermal conductivity of porous dust aggregates covering the ices. 

As cometary ices are predominantly made of water, with a fraction of volatile compounds, the morphology and evolution of water ice should be considered in order to understand their desorption. Several studies have highlighted the influence that the astrophysical environment has on the morphology of the ice, its porosity, the transition from an amorphous to a crystalline form, as well as how such characteristics are reflected in the spectral band of the ices in the mid-infrared (mid-IR) \citep{ dowell1960low,bergren1978oh,hagen1981infrared,laufer1987structure,jenniskens1997liquid, raut2007compaction, isokoski2014porosity, cazaux2015pore, bossa2015porosity}. A cometary nucleus model was developed that considers the effect of the phase change of water ice from amorphous to crystalline or clathrate forms on the outgassing of volatiles \citep{marboeuf2012cometary}. 

The thermal desorption of astrophysically relevant species on amorphous and crystalline water has been measured by several works, such as \citet{collings2004laboratory}, \citet{minissale2022thermal}, \citet{martin2014thermal}, and references therein. The diffusion of volatile compounds through amorphous water ice has been studied to determine the diffusion coefficient of these species \citep{he2018measurements}, as well as how the re-organization of water with temperature influences this diffusion \citep{mate2020diffusion}. Other experimental studies with similar objectives as \cite{may2013release, alan2013release} analyzed the release of trapped gases such as Ar, Kr, Xe, CH$_4$, N$_2$, O$_2$, and CO from amorphous solid water (ASW), using a crystallization-induced mechanism.

In the present work, an investigation of the desorption of a volatile compound (CH$_4$) through water ice during thermal processing has been performed. A methane fraction of 1\% with respect to water, as in \cite{mumma1996detection}, has been chosen. Temperature cycles from 30 K to 140 K were performed to mimic temperature cycles along the orbit of a comet and to address the volatile release before water sublimation. To mimic the effect of a dusty crust of comets, in some experiments, indene (C$_9$H$_8$), a less volatile species than water, is added on top of the ice mixtures. 
Our experimental setup does not allow us to make a layer of more realistic crust (such as dust), because non-volatile materials cannot be deposited in the experiments, and so we chose indene to study the effect of a covering layer composed of carbon and sublimating at a higher temperature than water.

This paper is structured as follows. In section \ref{1}, we describe the experimental setup and procedure for the experiments. In section \ref{chapres}, the results and the discussion are presented and divided into subsections where desorption of methane is studied in order to observe the effect of the variations of the heating ramp and type of ice deposition, temperature cycles, or presence of a crust on the upper layer. Finally,  in sections \ref{application} and \ref{concl}, we discuss the applications of our findings for comets and provide a summary of our conclusions, respectively.

\section{Experimental setup} \label{1}

\begin{figure}[!ht]
   \centering
   \includegraphics[width=8.5cm]{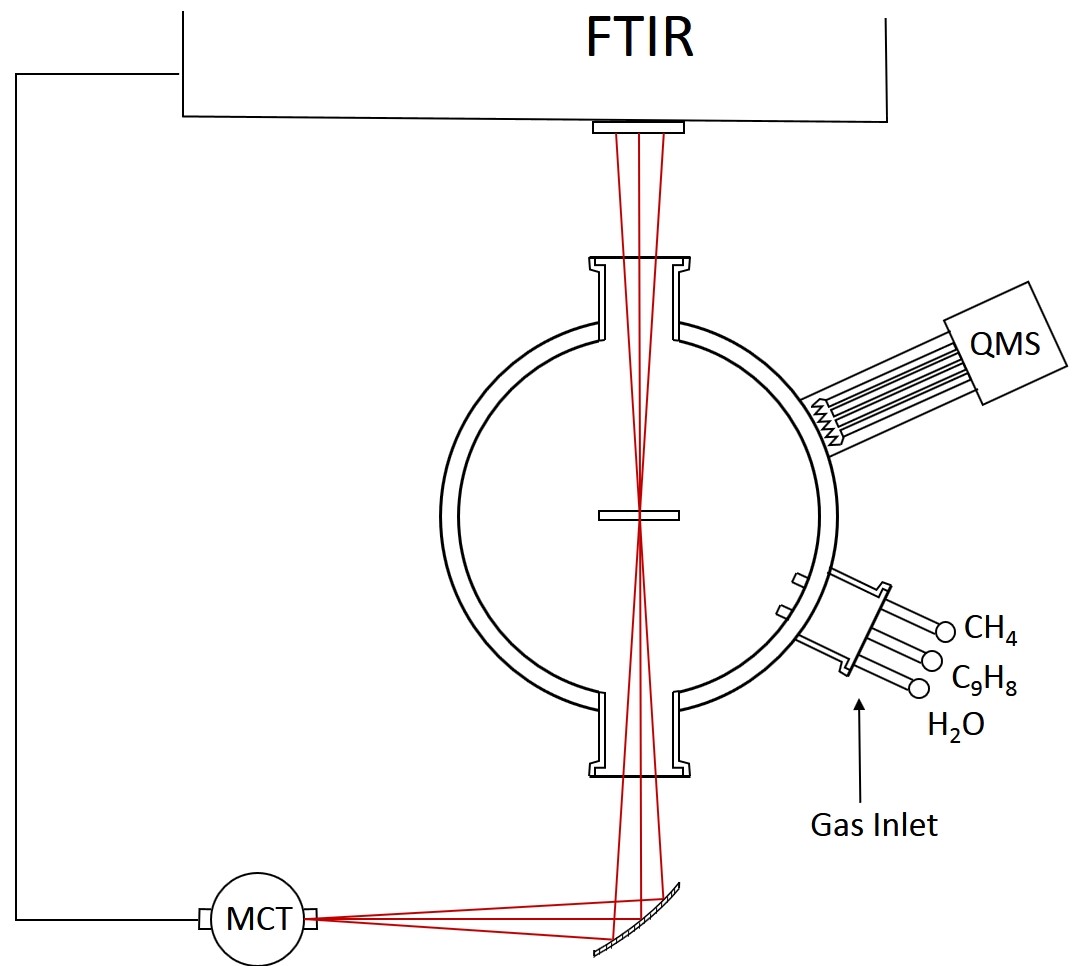}
      \caption{
              Scheme of the experimental setup. A transversal section of the cylindrical vacuum chamber is presented at the optical plane. The IR radiation enters the chamber through KBr windows. The inlet ports for the gas lines and the quadruple mass spectrometer are located in a lower plane. }
         \label{Figsetup}
   \end{figure}

The setup used here was previously described in \cite{mate20212}. Briefly, it consists of a high vacuum chamber provided with a cryostat (ARS DE-204AB) whose cold head can reach 10 K within 45 min. The cold head is provided with a heating foil, and its temperature can be controlled between 10 K and 300 K with an accuracy of 0.5 K using a Si diode and a Lakeshore temperature controller. A Cu sample holder designed to record IR transmission spectra of the sample is placed in close thermal contact with the cold head. A Si wafer of 25 mm in diameter and 1 mm in thickness is located on the holder leaving an exposed surface of 10 mm in diameter. The temperature is measured in the Cu sample holder, and indium foil is placed between the Cu holder and the Si wafer to improve the thermal contact. The background pressure in the chamber is in the range of 10$^{-8}$ mbar  at room temperature and can get to 10$^{-9}$ mbar with the cryostat turned on. The chamber is linked to a Bruker Vertex70 FTIR spectrometer through KBr windows. A Hiden HD200 quadrupole mass spectrometer directly connected to the chamber allows us to monitor the gas phase species during the deposition process and to perform temperature programmed desorption (TPD) experiments (see Figure \ref{Figsetup}).

Methane (99.95 $\%$, Air Liquide), distilled water (three times freeze-pump-thawed), and indene are introduced into the chamber through independent inlets. The methane line is equipped with a 2 sccm mass flow controller (Alicat) and the water and indene lines are controlled with leak valves.  C$_9$H$_8$ is a commercial liquid ($\ge$ 99 \%, Sigma-Aldrich) with 1 mbar vapor pressure at 20 °C. In the vacuum chamber, the inlet port of the three lines is located  10 cm below the cryostat's cold finger to ensure that the molecules that enter the chamber fill it with a homogeneous pressure (background deposition). The gases will condense forming two ice layers of equal thickness on the two sides of the IR transparent Si substrate.

Methane--water ice mixtures are grown at 30 K either by co-deposition, introducing both gases simultaneously into the vacuum chamber, or by sequential deposition, introducing first methane and then water. All the mixtures studied have a similar stoichiometry of about 1 \% CH$_4$ in water, and a thickness of about 750 nm/side (estimated assuming a density  for the mixture of pure water ice at 30 K, 0.64 g/cm3, taken from \citep{dohnalek2003deposition}). With the deposition pressures in the 10$^{-4}$ - 10$^{-5}$ mbar range employed, the ice layers were grown in less than 5 min. We performed TPD experiments of the mixtures  at 1 K/min and 5 K/min from 30 K to 200 K. Also, TPDs of pure methane ices grown at 30 K were carried out for comparison.  The heating ramp of 5 K/min is close to the fastest ramp provided by the setup, and the 1 K/min ramp is chosen as the slowest one that allows us to perform reliable TPD experiments. Due to the background pressure in the chamber, a water ice layer of 13 ML will grow during the 1 K/min ramp, which is not expected to affect the results obtained in this work. At lower rates, the contamination of the ices due to the background pressure in the high-vacuum chamber could become significant. In another set of experiments, similar CH$_4$:H$_2$O ice mixtures were thermally processed with temperature cycles (warm up and cool down). The samples were warmed to 140 K, cooled down to 30 K, warmed up again to 140 K, and then finally cooled back to the initial temperature of 30 K. The temperature cycles are performed without stopping at the higher temperature (140 K) and using the same heating and cooling ramps. For the two cycles experiments, the ASW layer grown on top due to background pressure will be about 50 ML in thickness and is still is not expected to affect the interpretation of the results. In a last set of experiments, to simulate comet nuclei crust, indene layers were grown on top of the CH$_4$:H$_2$O ice mixtures. TPD experiments, at rates equal to the ones performed on the samples without crust, were carried out for crust-covered ices. Indene layers of different thicknesses were grown to study the effect of the top crust on methane desorption curves. 

In the TPD experiments, mass 16 was chosen to monitor the sublimated CH$_4$ molecules, and its time evolution was followed with the fast scan mode using electron multiplier detection and sampling every 4 seconds. Infrared spectra of the ices were taken before the TPD and at controlled time intervals during the experiments. The spectra were recorded in normal transmission configuration, with 4 cm$^{-1}$ resolution, and the accumulated scans and the time interval between spectra varied depending on the heating ramp used. For ramps at 5 K/min, spectra were taken every minute, accumulating 100 scans. For the slower ramp, the spectra were taken at 5 min intervals, accumulating 300 scans.

The column density of the molecules present in the ices grown was derived from the IR absorbance spectra employing the IR band strengths available in the literature. The OH-stretching (symmetric and asymmetric) band at 3200 cm$^{-1}$ for water and the v$_4$ bending mode at 1300 cm$^{-1}$ for methane were considered. For indene, the most intense band in its mid-IR spectrum was employed, which appears at 760 cm$^{-1}$. The IR absorption coefficients for these bands are: A$_{H_2O}$$_{3200}$ = 2$\times$10$^{-16}$ cm molec$^{-1}$ \citep{mastrapa2009optical} (we note that the latter does not take into account ASW density variations with temperature), A$_{CH_4}$$_{1300}$ = 7$\times$10$^{-18}$ cm molec$^{-1}$ \citep{molpeceres2017physical}, and A$_{C_9H_8}$$_{760}$ = 2$\times$10$^{-17}$ cm molec$^{-1}$ \citep{submittedBelen}. 

Moreover, assuming the ice density is known, it is possible to obtain the  thickness of the ice layer. Values of $\rho_{CH_4}$ = 0.46 g/cm$^{3}$ \citep{molpeceres2017physical} and $\rho_{H_2O}$ = 0.65 g/cm$^{3}$ \citep{dohnalek2003deposition} have been measured for pure CH$_4$ and H$_2$O ices at 30 K, respectively. For co-deposited ices, the thickness of a pure species layer containing the same number of molecules of CH$_4$ or H$_2$O  is considered. To our knowledge, there is no published value for the density of amorphous indene ice, and so the density of the liquid will be assumed in this paper to make an estimation of the thickness of the indene ice layers grown. Indene liquid density is $\rho_{C_9H_8}$ = 0.997 g/cm$^{3}$ \citep{linstrom2001nist}. 

The estimated thickness of the ice layer for the ices grown in this work is about 10 and 600 nm for methane and water, respectively. Indene layer thickness has been varied between 25 nm and 150 nm.

\section{Results and Discussion}\label{chapres}

\begin{figure}[!ht]
   \centering
   \includegraphics[width=9.5cm]{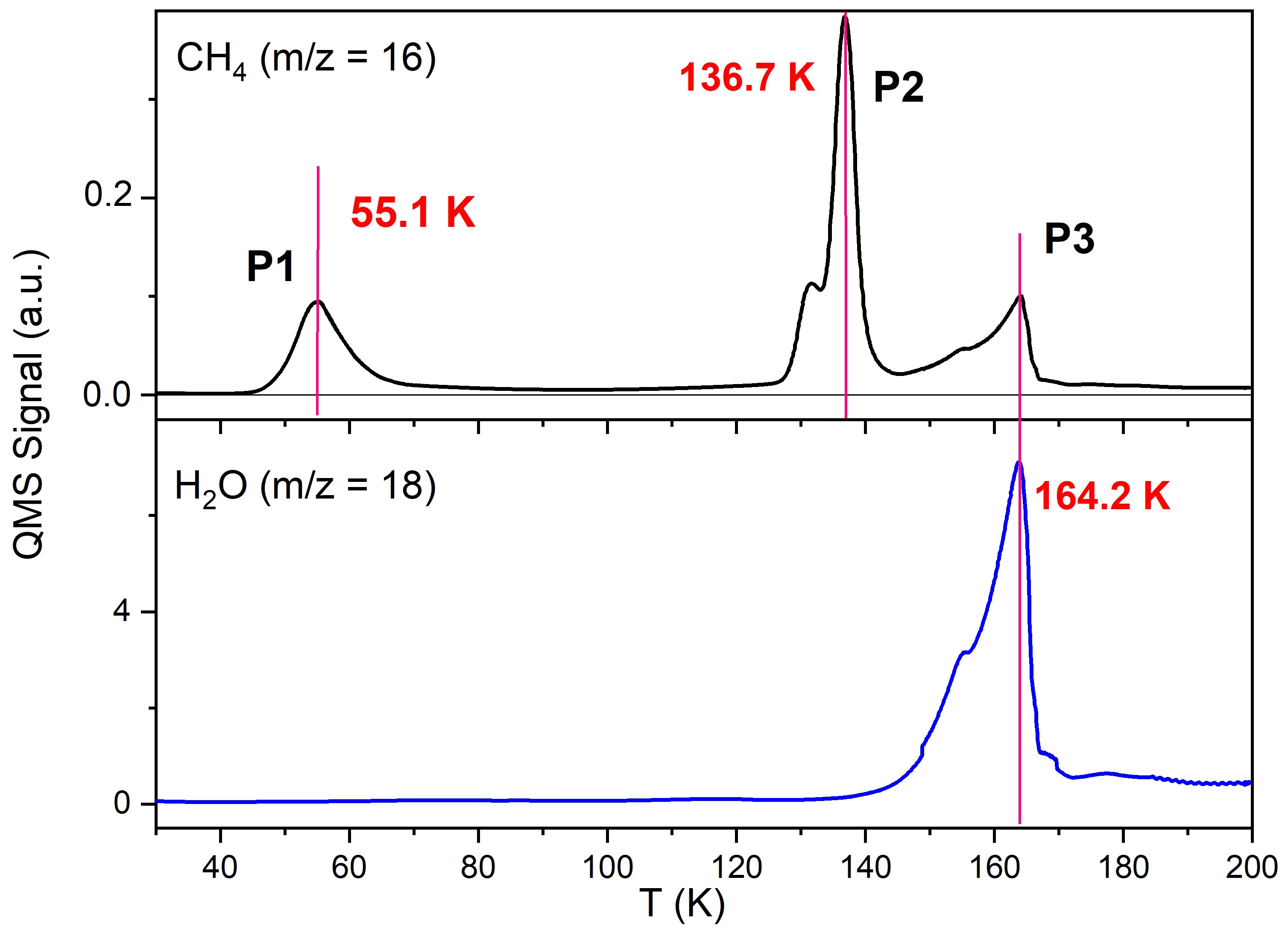}
      \caption{TPD curves of CH$_4$ (top panel) and H$_2$O (bottom panel) for a CH$_4$:H$_2$O layered ice heated at 1 K/min. 
              }
         \label{simpledes}
   \end{figure}

When a TPD of an ice mixture of methane and water is performed from 30 K to 200 K, three desorption peaks (as also reported in \cite{collings2004laboratory}) appear for methane: one around 50 K (P$_1$), one around 140 K (P$_2$), and another one between 160 and 180 K (P$_3$), while only one peak appears for water, at around 170 K (Figure \ref{simpledes}).

We provide the following explanations for the three peaks for methane: 

\begin{enumerate}
    \item P$_1$: The sublimation temperature of weak interacting methane (with water) is 60 K. This peak, although affected by water ice porosity and the binding and diffusion energies of CH$_4$, is mainly caused by the transition from high- to low-density amorphous solid water that takes place between 40 and 70 K \citep{jenniskens1994structural}. The relation between the low-temperature peak and the high-to-low-density water ice transition is claimed for the first time in this work, but is supported by the experiments performed by \cite{ayotte2001effect}. These authors studied the desorption of various gases (N$_2$, Ar, O$_2$, CO, and CH$_4$) through porous ASW and observed a low-temperature desorption peak that occurs at about the same temperature (below 80 K), independently of the gas species. This observation proves that P$_1$ desorption is led by the  restructuring of water ice, which is mentioned by \cite{ayotte2001effect}. However, because porous reorganization is a continuous process, the ice density phase transition in the 40-70 K interval must play a key role in the appearance of a desorption peak in that temperature range.
    \item P$_2$: Crystallization of water takes place at around 140 K. During water-ice transformation from amorphous to crystalline, methane is pushed out of the water matrix. This peak is known as volcano desorption.
    \item P$_3$: Co-desorption of methane with water takes place between 160 and 180 K. 
\end{enumerate}

It is interesting to notice that methane continuously desorbs in between peaks P$_1$ and P$_2$, as indicated by the QMS signal being above zero in that temperature interval (see top panel of Figure \ref{simpledes}). This continuous desorption is due to the reorganization of amorphous water ice, which causes a change in the size and shape of the porous structure (pores coalesce) upon heating \citep{cazaux2015pore}, as can be seen in the IR spectrum of water (see section \ref{tempcycsecwat}). The reorganization allows methane molecules to diffuse through the pores and to sublimate continuously between 50 and 140 K. Additionally, the amorphous to crystalline water-ice phase transition that occurs in P$_2$ is not complete (see section \ref{tempcycsecwat}), and a small fraction of water ice remains in its amorphous form. This fraction is correlated with the number of impurities that are retained and released during P$_3$. 
 
\subsection{Effect of the heating ramp and type of deposition on the desorption of methane through water ice} \label{effectheatingramp}

\begin{figure}[!ht]
   \centering
   \includegraphics[width=9cm]{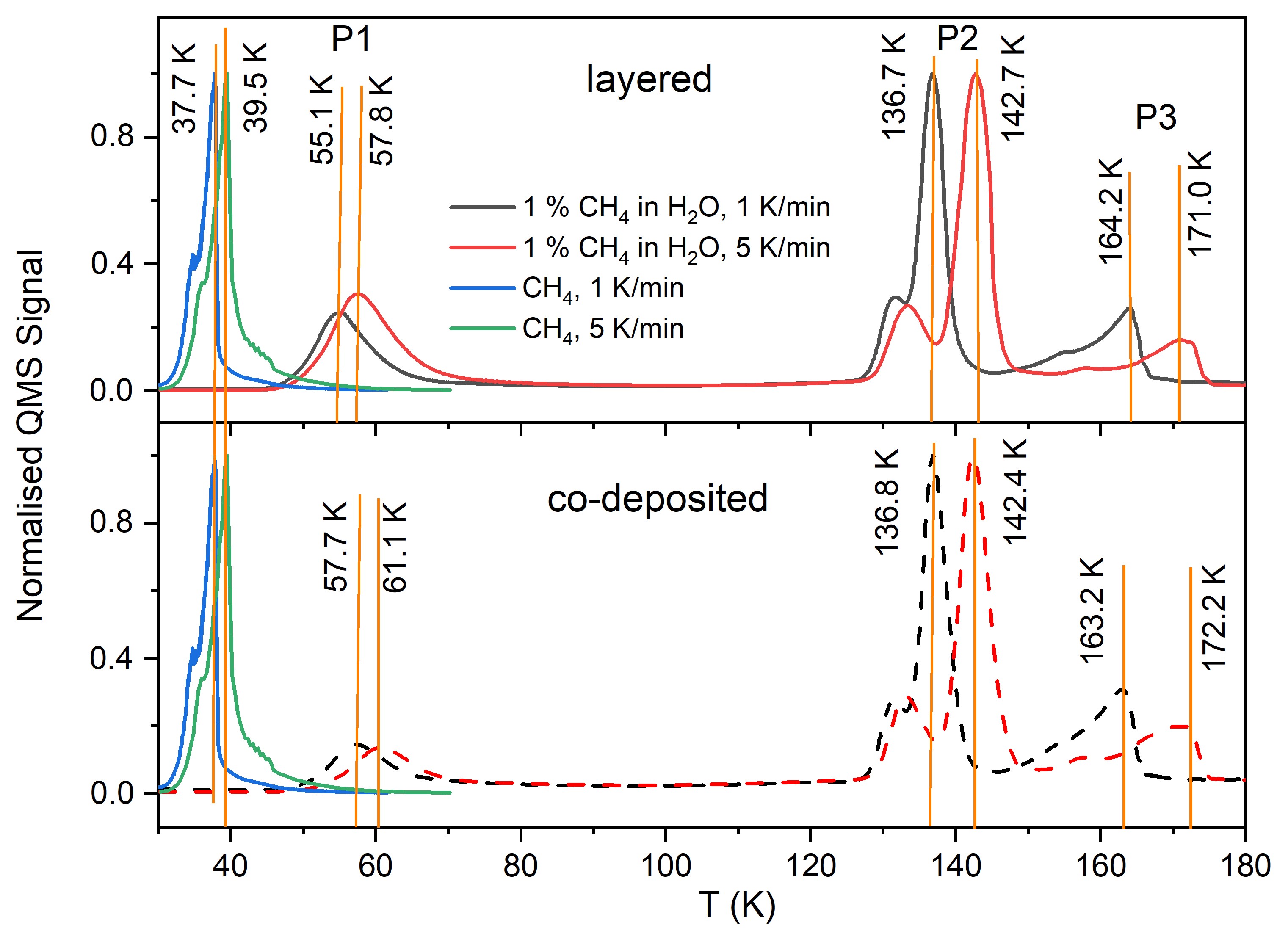}
      \caption{TPD curves of CH$_4$, normalized to P$_2$, for different heating ramps and H$_2$O:CH$_4$ ice mixture morphologies. TPDs of pure CH$_4$ ices are also shown, in green and blue, in both panels for reference. Dashed lines indicate co-deposited cases and solid lines indicate layered ones. Black lines are for a 1 K/min heating ramp, and red ones are for a 5 K/min heating ramp. } 
         \label{laylaymixmix}
   \end{figure}

The effects of the variation of both the warming ramp and the type of deposition (layered or co-deposited) on the desorption of methane through water ice are seen in Figure \ref{laylaymixmix}. It is observed that, the faster the ramp, the higher the desorption peak temperature. This is a well-known effect that can be described with the Polanyi-Wigner model of thermal desorption from surfaces, as illustrated for example in the simulations by \cite{brown2007fundamental}. A similar effect in P$_1$, P$_2$, and P$_3$ is observed in layered and co-deposited ice mixtures, with an increase of around 3 K in P$_1$ desorption temperature, around 6 K in P$_2$ desorption temperature, and 7 to 9 K in P$_3$ due to a 1 to 5 K/min heating ramp increase.

It is also observed that the heating ramp affects the P$_1$ peak intensity in the layered case. In the top panel of Figure \ref{laylaymixmix}, it can be seen that P$_1$ is larger in the 5 K/min ramp than in the 1 K/min ramp. We propose a possible cause for this behavior. A faster heating ramp implies a large heat absorbed by the lower layer of CH$_4$ molecules in less time, and given that diffusion through the ASW layer is not instantaneous, a larger fraction of CH$_4$ molecules at a temperature above sublimation are present in the lower layers of ASW. Consequently, as CH$_4$:H$_2$O interaction is stronger than CH$_4$:CH$_4$ interaction, CH$_4$ molecules are pushed or dragged through the pores of the ASW layers more efficiently, diminishing the number of CH$_4$ molecules that remain trapped.

\begin{figure}[!ht]
   \centering
   \includegraphics[width=9cm]{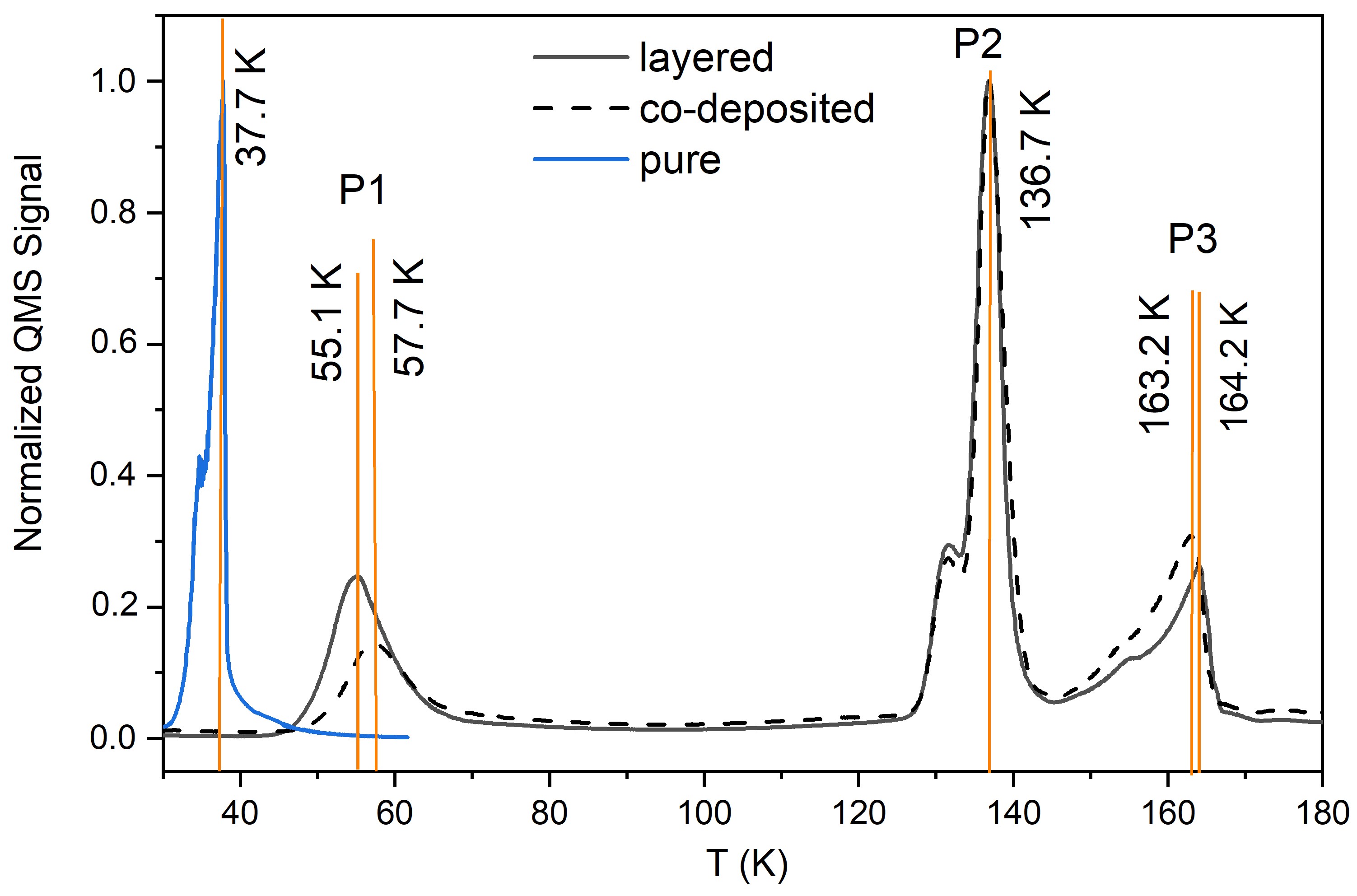}
      \caption{Methane TPD curves performed at 1 K/min normalized to P$_2$, for different deposition types of CH$_4$:H$_2$O ice mixtures. 
              }
         \label{laymix}
\end{figure}

In order to showcase the differences in the TPDs associated with the different methane--water structures, we plot the 1 K/min experiments (shown in Figure \ref{laylaymixmix}) together  in Figure \ref{laymix}. The greatest effect of the ice mixture structure is seen in P$_1$, which changes in intensity and peak position. This peak appears more intense and at around 55 K in the layered ice mixture, while it is weaker and peaks at a 2.4 K higher temperature in the co-deposited ice. To understand the different intensity of P$_1$ in layered versus co-deposited samples,  as explained above, we must remember that for layered ices, CH$_4$ sublimates and travels through the porous structure of ASW. However, the desorption of CH$_4$ in the co-deposited ices can be understood differently. As CH$_4$ is mixed with water, CH$_4$ sublimation is reached at higher temperatures because its binding energy with water is higher than with itself. In this case, CH$_4$ diffuses through pores less efficiently than in layered ices, and the desorption is governed by the reorganization of water ice.

In contrast with P$_1$, the positions of P$_2$ and P$_3$ are not significantly affected by the ice mixture structure. This is because P$_2$ is related to the crystallization of the water ice, and P$_3$ to the water sublimation temperature, and these are not markedly affected by the small fraction of CH$_4$ (around 1 \%) present in the ice mixtures.

The binding energies of molecules in the ices are directly linked to their temperature of desorption \citep{minissale2022thermal}. It is known that CH$_4$:H$_2$O interaction is stronger than  CH$_4$:CH$_4$ interaction \citep{luna2014new,he2016binding,smith2016desorption}. As there is a larger number of CH$_4$ molecules interacting with water molecules in co-deposited ices than in the layered ones, this could be one of the explanations for why methane P$_1$ desorption occurs at higher temperatures in those mixtures. 

The ice morphology is also able to explain the larger amount of CH$_4$ molecules released and the lower temperature of P$_1$ in the layered sample. In this case, the methane molecules can more easily diffuse through the routes made by the pores in the top  layer of water ice, because they were initially in a lower layer and are not trapped in the ice matrix as in the co-deposited case. Complementary to this, in P$_3$, the co-deposited case shows a larger desorption, which is due to the larger amount of methane that has remained trapped in this mixture. In Table \ref{table1}, the peak area ratios (P$_1$/P$_2$ and P$_3$/P$_2$) are listed for the experiments presented in Figure \ref{laylaymixmix}. They are all less than 1, showing that the highest quantity of methane is released in P$_2$ during water crystallization. It can be seen that comparing the experiments performed with the same heating ramp, P$_1$/P$_2$ is larger for the layered case than for the co-deposited one, while consequently P$_3$/P$_2$ shows the opposite behaviour.

   \begin{table}[!ht]
   
      \caption[]{Methane desorption peak area ratios for the TPD experiments shown in Figure \ref{laylaymixmix}.}
      \label{table1}
      \centering
      \scriptsize
      \scalebox{1.3}{
         \begin{tabular}{ccccc} \hline
            \multirow{2}{*}{Ratio} & Layered & Co-deposited & Layered & Co-deposited \\
            & 1 K/min & 1 K/min & 5 K/min & 5 K/min
            \\ 
         
            \hline
            
            P$_1$/P$_2$ & 0.387 & 0.249 & 0.455 & 0.207   \\
         
            P$_3$/P$_2$ & 0.396 & 0.487 & 0.321 & 0.434 \\
            \hline
         \end{tabular}}
   \end{table}

\begin{figure}[!ht]
   \centering
   \includegraphics[width=8.5cm]{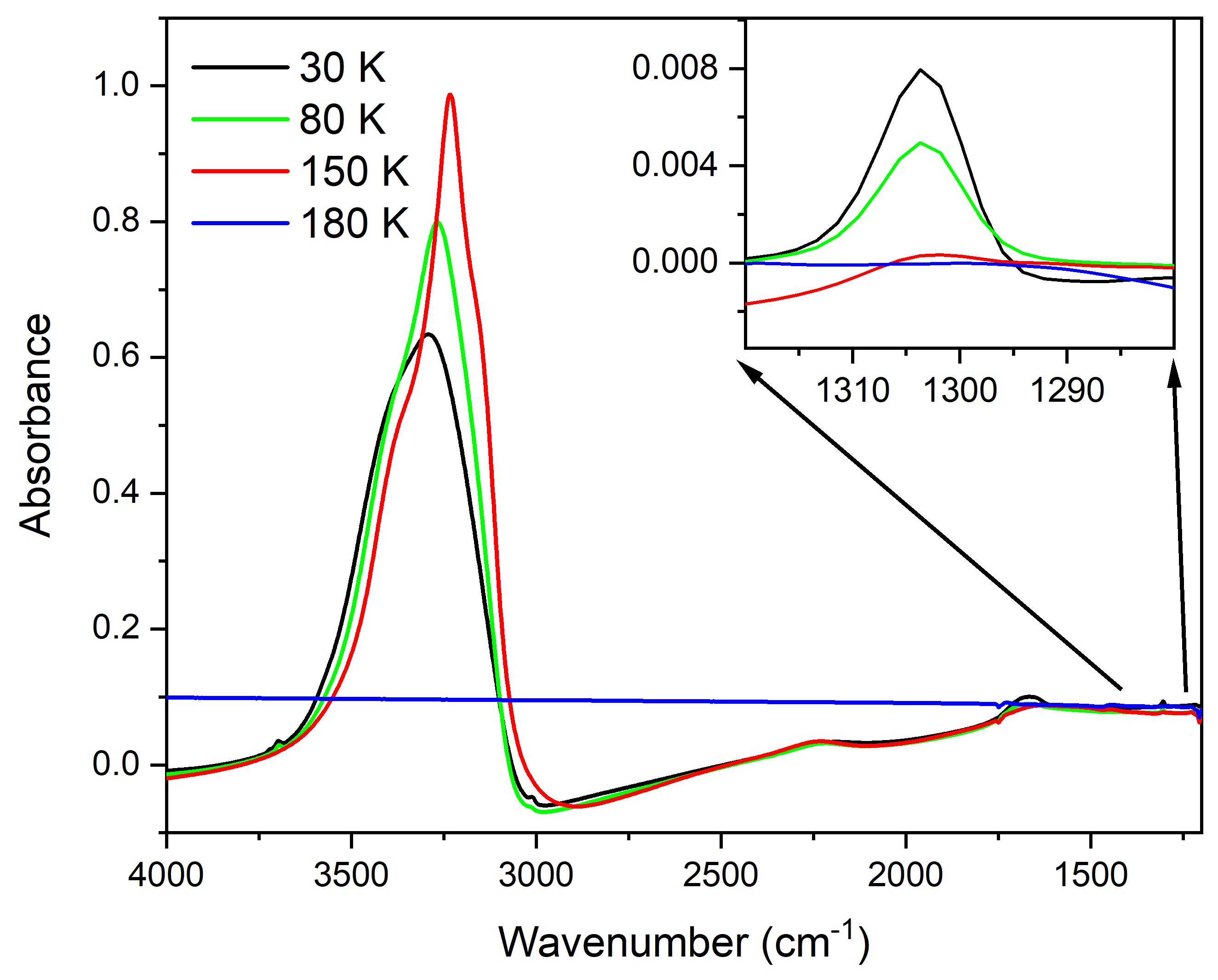}
      \caption{Infrared spectra from 3800 cm$^{-1}$ to 1200 cm$^{-1}$ (wavenumber) for the co-deposited experiment with a 1 \% ratio stoichiometry. The material is heated to temperatures of 30, 80, 150 and 180 K  with a heating ramp of 1 K/min. 
              }
         \label{IRspecallexp}
\end{figure}

Figure \ref{IRspecallexp} shows the IR absorbance spectra of a co-deposited mixture of water and methane heated  to four different temperatures (30, 80, 150, and 180 K) with a ramp of 1 K/min. The first temperature corresponds to the deposition temperature (30 K), and the others to the end of each of the three methane desorption peak temperatures (80, 150, and 180 K, respectively). The shape and intensity changes in the spectra show not only the internal changes in the amorphous water structure but also the phase transition from amorphous to crystalline. The insert within Figure \ref{IRspecallexp} highlights the 1300 cm$^{-1}$ band of CH$_4$ as it follows the methane loss during the heating process, which is consistent with that observed in the TPD experiments and quantified in Table \ref{table1}. However, methane does not desorb only at these three specific temperatures but in a more continuous way throughout the TPD experiment, which is due to the continuous change in ASW structures as the material warms up \citep{jenniskens1994structural}. A more detailed analysis of the water structural changes is presented in section \ref{tempcycsecwat}. 

\subsection{Temperature cycles} \label{tempcycsecwat}

In this section, we grow the same ices as in section \ref{effectheatingramp}, but this time we carry out two temperature cycles, meaning that the temperature is increased until 140 K, decreased to 30 K, increased again to 140 K, and then decreased again to 30 K. Such cycles are meant to mimic the thermal evolution of comets during several orbits around the Sun and to observe their effects on the desorption of volatile compounds. This type of experiment is mainly used to study the behavior of the internal layers of the nucleus of the comet, which do not reach 200 K even at perihelion and therefore do not immediately desorb from the nucleus of the comet, as the external layer does. 

\begin{figure}[!ht]
   \centering
   \includegraphics[width=11cm]{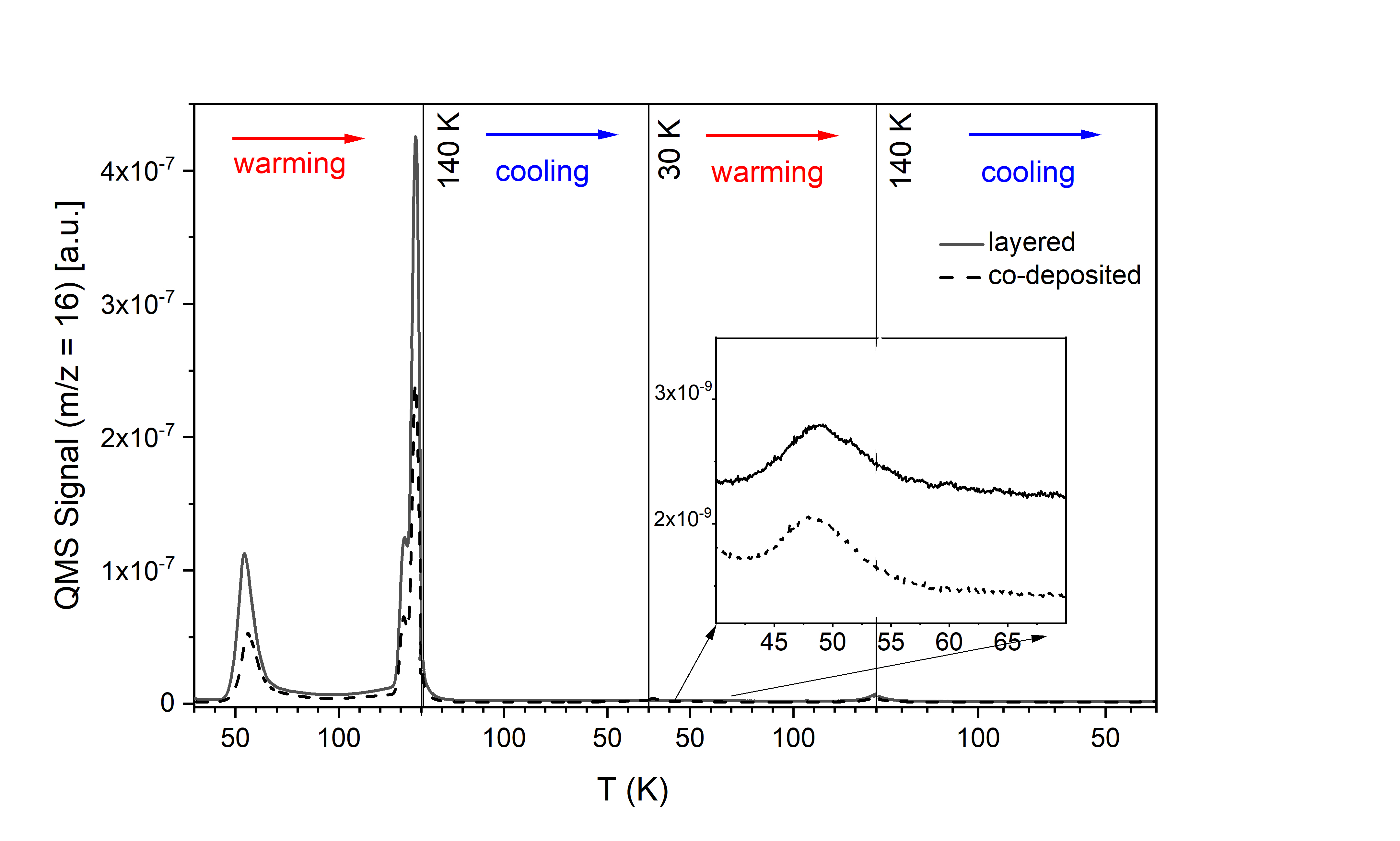}
      \caption{Methane TPD curves for H$_2$O:CH$_4$ ice mixtures during two temperature cycles. 
              }
         \label{cycles}
   \end{figure}

 Figure \ref{cycles} shows the methane TPD of two 30-140-30-140 K heating-cooling cycles at 1 K/min for layered and co-deposited CH$_4$:H$_2$O ice mixtures. The temperature on the x-axis increases and decreases depending on whether it is a heating or a cooling process. The vertical lines indicate the change from heating to cooling or vice versa. After one cycle of heating and cooling, some methane is still retained in the water ice matrix, and in the second cycle of heating a P$_1$ desorption peak is still detectable (see insert in the figure).

For the first cycle, we obtain the same results as in Figure \ref{laymix} with a shift in temperature for P$_1$ of about 2 K between layered and co-deposited. During the second cycle, for both structures P$_1$ and P$_2$ show their maximum at lower temperatures than in the first cycle. This can be explained by two main factors. First, in the second cycle, these temperature shifts in the peaks  are a result of the smaller amount of methane trapped in the water ice. Secondly, they reflect the difference in the capacity of methane to diffuse through amorphous versus crystalline water ice. When water molecules reorganize themselves in the crystallization process, they push CH$_4$ out of their structure, reducing the van der Waals interactions. Therefore, the CH$_4$ that is retained in the water ice after the first cycle will be able to sublimate faster in the second one \citep{mate2020diffusion}. Looking at the difference between layered and co-deposited samples, although during the first cycle,  P$_1$ is different for each case, as is discussed in the previous section, during the second cycle the morphological differences are reduced and P$_1$ appears similar. However, it is still shown that the amount of methane released in the second cycle for the layered sample is a little higher than in the co-deposited case (as seen in Table \ref{table3}).

Table \ref{table3} reports the ratio CH$_4$/H$_2$O of the number of molecules in the ice at the end of both the first and the second cycle. The number of molecules of CH$_4$ or H$_2$O present in the ice was estimated from the IR spectra using the procedure described in section \ref{1}. At the end of the first cycle, considering that the ratio between methane and water at the beginning of the cycles is almost the same, a more important fraction of methane is retained for the co-deposited case. At the end of the second cycle, for the layered case, methane is almost fully desorbed (around 95 \%). Instead, for the co-deposited case, a small fraction (around 15 \%) remains in the ice.

   \begin{table}[!ht]
      \caption[]{CH$_4$/ H$_2$O   molecule ratios in the ice mixture at 30 K, after generation, and after the first and the second heating-cooling cycles.}
         \label{table3}
         \centering
         \scriptsize
      \scalebox{1.4}{
         \begin{tabular}{ccc}
            \hline
            Cycles & Ratio layered & Ratio co-deposited \\
            \hline
            
           beginning & 1.3 \% & 1.1 \% \\
         
           end 1st & 0.14 \% & 0.25 \% \\
          end 2nd & 0.07 \%  & 0.16 \% \\
           
            \hline
         \end{tabular}}
   \end{table}   

Methane is more intrinsically mixed with water in co-deposited ices and disturbs water crystallization. Although the authors performed no calculations as to the effect of several approaches close to the Sun, \cite{marboeuf2012cometary} did find that mixed ices release trapped volatile compounds  less efficiently, which is in agreement with our experiments.

\begin{figure}[!ht]
   \centering
   \includegraphics[width=9cm]{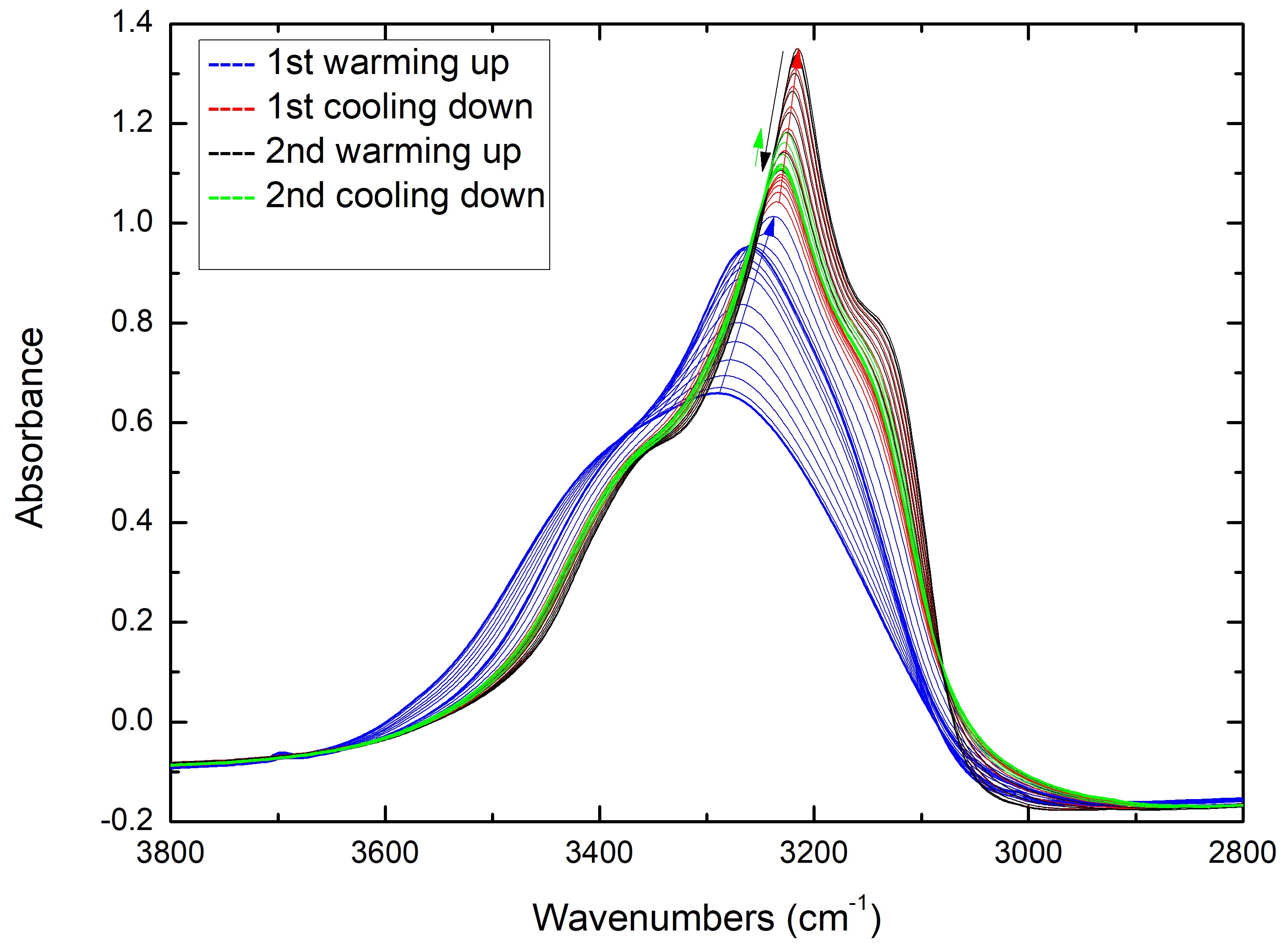}
      \caption{Infrared spectra of water during the entire duration of two temperature cycles. The arrows indicate the direction of evolution.
              }
         \label{spectrawat}
   \end{figure}

Water is deposited at 30 K and is therefore amorphous in its high-intrinsic-density form \citep{jenniskens1994structural}. As it warms up to 140 K, amorphous to partially cubic crystalline transformation takes place; see \citet{hagen1981infrared} and \citet{jenniskens1994structural}. The evolution of the water ice spectra during the two warming-cooling cycles is presented in Figure \ref{spectrawat}, where the spectra corresponding to each different ramp value are represented in different colors. The change of phase during the first heating ramp is clearly seen in the blue spectra. For the other three ramps (first cooling, second warming, and second cooling, in red, gray, and green,  respectively), the shape of the spectrum does slightly vary from red to green, through gray, due to the molecules reorienting themselves. After the first warm up, all the spectral changes observed in Figure \ref{spectrawat} are explained by reversible changes due to OH-stretching vibrations of the intermolecularly coupled OH oscillators, as explained by \cite{bergren1978oh} and \cite{hagen1981infrared}. This can be seen through the overlap between the red, gray, and green spectra. The greatest changes take place during the first warming because energy is transferred to the ice, causing the molecules to reorganize, which leads to irreversible changes that eject most of the methane (as seen in Table \ref{table3}). During the other periods of cooling and heating in the cycles, no appreciable variations are noticed. 

During the reorganization of the molecules in the ices mentioned above, water molecules diffuse until they are more stable (they increase their binding energies by having more neighbors), which subsequently allows the pores to merge and to make channels for methane to diffuse \citep{cazaux2015pore}. After crossing 140~K, the lack of variation in the IR spectra suggests that water-ice reorganization no longer occurs. This implies that the desorption of methane, which is a by-product of the reorganization between 50 and 140~K (as shown in Figure \ref{simpledes}), only occurs when amorphous water ice is heated to 140~K for the first time. 

\subsection{Indene (crust) layer on top of the CH$_4$:H$_2$O mixture}

\begin{figure}[!ht]
   \centering
   \includegraphics[width=9cm]{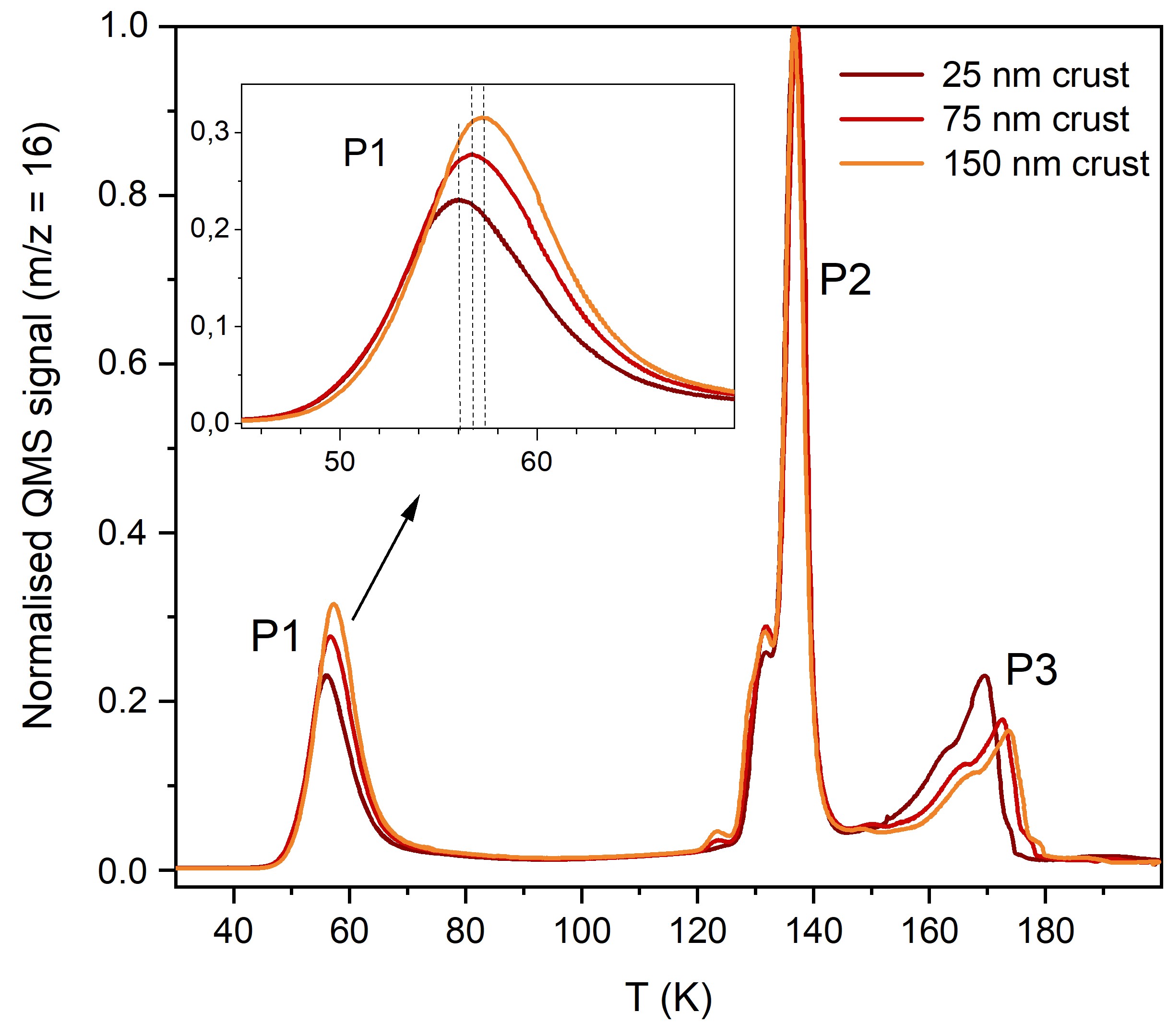}
      \caption{Normalized QMS signal of methane as a function of temperature for different thicknesses of the indene crust on top of the CH$_4$:H$_2$O mixture.
              }
         \label{crust}
   \end{figure}

Figure \ref{crust} presents the TPD experiments performed with a layer of indene grown on top of the CH$_4$:H$_2$O mixture. These experiments were performed with only co-deposited mixtures in order to get an idea of the effect of the refractory layer of CH$_4$ desorption. When adding a layer of indene, P$_1$ is shifted to higher temperatures, and 
the thicker the indene layer, the greater the shift (see insert in Figure \ref{crust}). These temperature shifts can be caused by the longer times that CH$_4$ molecules need to travel through thicker indene crusts before being released. Longer times imply a release at higher temperatures during the TPD experiments. 
A similar behavior was observed in previous experiments that studied the outgassing of water ice through dust layers of different thicknesses \citep{gundlach2011outgassing}. 

\begin{figure}[!ht]
   \centering
   \includegraphics[width=9cm]{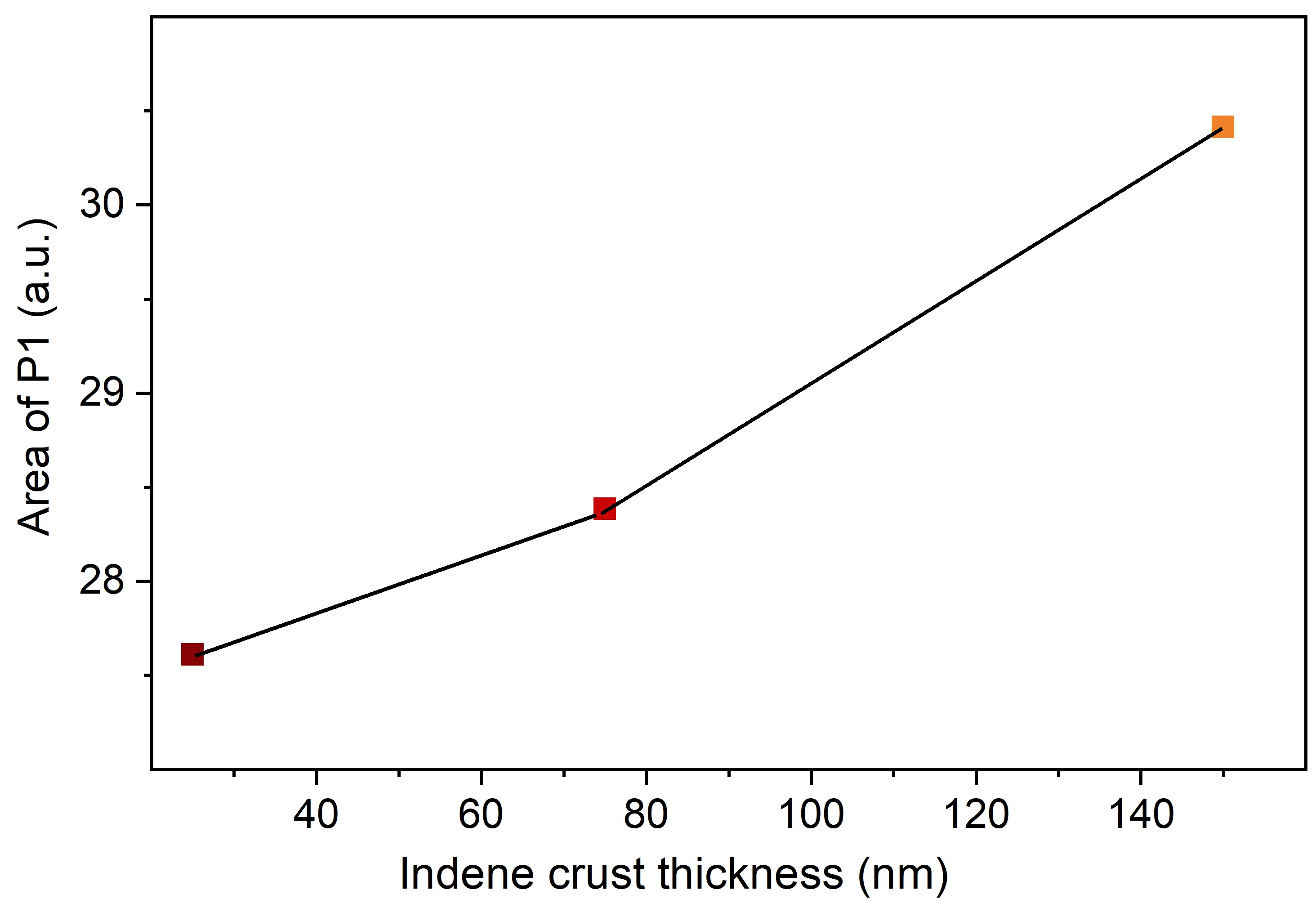}
      \caption{ Area of the P$_1$ desorption peak of methane versus indene layer thickness.
              }
         \label{trespuntosindene}
   \end{figure}

Interestingly, the amount of CH$_4$ desorbed in P$_1$ increases with the thickness of the crust (see  Figure \ref{trespuntosindene}). In our experiments, the volatile CH$_4$ has to diffuse through both the water ice and the indene crust top layer. A possible explanation for the behavior observed, which is an increase in the amount of 
CH$_4$ released with increasing crust thickness, is based on the reorganization of the water-ice matrix during TPD combined with the insulating effect of the indene crust. First, it is assumed that the indene layer is a resistor to gas flow, the diffusion of CH$_4$ molecules through it being slower than through water ice. In this scenario, during the TPD, when CH$_4$ molecules are trying to escape through the ASW, the presence of the crust forces some of them to stay for a longer time in the ASW ice (we remind the reader that gas molecules move randomly), causing an increase in the CH$_4$ “pressure” in the ice. The higher the number of CH$_4$ molecules, the fewer the interactions they have with H$_2$O within the ice, and fewer are retained in its structure. Finally, the increase in CH$_4$ pressure within the ice will be more pronounced for thicker indene layers, and, as higher pressure leads to less retention in the ice, more CH$_4$ will be liberated.

Another cause could be that it takes longer to grow thicker indene layers and, during that extra time at 30 K, water ice can restructure and reorganize, leading to the coalescence of pores (\cite{bossa2015porosity}). Additionally, during the deposition of indene, the indene kinetic energy can in part reach the water ice and be employed in reorganization. Any extra water reorganization will increase the CH$_4$ release from its structure.

For P$_3$, which is related to methane co-desorption with water, it is clear that, the thicker the crust, the lower the quantity of methane that desorbs, and this occurs at a higher temperature. This is because the CH$_4$ desorption observed in P$_1$ increases with the increasing thickness of the indene, implying that the CH$_4$ fraction remaining in the ice decreases with the increasing thickness of the indene. This CH$_4$ remaining in the water ice co-desorbs with water in P$_3$, releasing the CH$_4$ that did not desorb in P$_1$.

\section{Application for comets} \label{application}

In observations of the release of volatile compounds by comets, the desorption of species such as CO, H$_2$CO, and HCN increases with decreasing distance from the Sun. These species show desorption patterns, that is, slopes, that appear to increase from 7 to 3 au, then stagnate between 3 and 2 au, and increase again with a steeper slope between 2 and 1 au \citep{biver20021995}. In particular, CH$_4$ has been observed in the coma of several comets \citep{2002ESASP.500..753M, bockelee2004composition, bockelee2017composition}, but its low abundance precludes the representation of its evolution with heliocentric distance. Our experimental results show that the desorption of volatile compounds (that do not make hydrogen bonds) is governed by the reorganization of amorphous water. This suggests that for comets, before perihelion, the desorption of CH$_4$ would follow a similar slope to those of volatile compounds such as CO. The slope of the desorption of methane with decreasing and increasing distances to perihelion has been modeled for amorphous and crystalline ices, as well as clathrates, and results show that models considering ice in an amorphous state can produce desorption slopes that differ before and after perihelion \citep{marboeuf2012cometary}.

For CO in particular, looking at the desorption with distance for the comet Hale Bopp \citep{biver20021995}, the amount of CO increases by a factor 5.6 from 7 to 3 au, and by a factor 10 from 1.8 to 1 au. Such an increase can be compared to an increase in the desorption of molecules. The desorption of molecules with a binding energy E$_{bin}$ at a temperature T$_0$ can be computed as 

\begin{equation}
    R_{des}=\nu \times \exp{\frac{-E_{bin}}{T_0}}
.\end{equation}

The increase between distance 1 and distance 2, at which the temperature is T$_1$ and T$_2$, respectively, can then be written as
\begin{equation}\label{eq 3}
    I=\frac{\exp{\frac{-E_{bin}}{T_1}-\exp{\frac{-E_{bin}}{T_2}}}}{\exp{\frac{-E_{bin}}{T_1}}}
.\end{equation}

Between 7 and 3 au, molecules are desorbing. The temperature of the cometary surface at these two distances is assumed to be 130 and 220 K, respectively, taken from Figure 5a of \cite{prialnick2004}. To reproduce the increase in the amount of a volatile species by a factor of 5.6 between 7 and 3 au, its binding energy has to be $\sim$ 650 K (from equation \ref{eq 3}). This binding energy reflects the mobility of water molecules in the water ice during reorganization. The binding energy of water is $\sim$ 2200 K per hydrogen bound \citep{dartois2013}. For water molecules to be mobile in the ice, the barrier for diffusion is a fraction of the binding energy ($\sim$ 20 $\%$), which means that a water molecule bound in the ice with two hydrogens bound will have a diffusion barrier of $\sim$ 900 K. This is in agreement with the energy determined of about $\sim$ 800 K from the desorption slopes of Hale Bopp. Between 1.8 and 1 au, assuming temperatures of 280 and 365 K, respectively, from \cite{prialnick2004}, the binding energy to account for the increase in the desorption is of the order of 2900 K. This higher binding energy illustrates the regime in which water is sublimating and volatile species are co-desorbing. However, this binding energy is low compared to the binding energy of water ice (\cite{minissale2022thermal} and references therein). This suggests that this process is not completely driven by the desorption of pure water ice but that a large amount of dust also influences this process \citep{fulle2010comet}.

The restructuring of water ice is very sensitive to the temperature at which ices have been deposited. The fact that we see the desorption of volatile compounds due to reorganization of the water structure shows that the cometary material that is releasing the volatile compounds is processed for the first time –-which means that this material is pristine. This also indicates that such material has been deposited at low temperatures, allowing the existence of water ices mixed with CO or CH$_4$. This is in agreement with the cometary ices being formed in the pre-stellar cloud at very low temperatures \citep{altwegg2022}.

\section{Conclusions} \label{concl}

We studied the desorption characteristics of methane through amorphous solid water and a layer of indene (mimicking a crust) with thermal programmed desorption (TPD) experiments using mass and IR spectroscopy. We considered an abundance ratio of methane to water of 0.01 ---the typical ratio observed for cometary nuclei---, a heating ramp that allows us to study the evolution of the ices, and a deposition type that alternates between layered and co-deposited. The main conclusions of our work can be summarized as follows.

\begin{enumerate}
    \item During the TPD of an ice mixture of methane and water from 30 K to 200 K, three desorption peaks appear for methane: one around 50 K mostly due to ASW high-to-low-density phase transition; one around 140 K due to water ice crystallization; and another between 160 and 180 K due to co-desorption of methane and water ice. 
    \item In between the desorption peaks of methane, from 50 to 140~K, methane desorbs continuously but at a lower rate. This desorption is observed for the first time and is due to the reorganization of water ice.
     \item We performed two temperature cycles (warming up from 30 to 140 K, cooling from 140 to 30 K) and observed that water ice morphology no longer changes after passing 140~K. This means that the continuous desorption of methane, observed in our experiments during the first warming, occurs only during the reorganization of water. 
     \item The higher the thickness of a layer of crust on top of the mixture, the higher the temperature at which methane desorbs, as molecules trapped in the ices have more difficulty desorbing when a crust is present. Therefore, more orbital cycles will be needed for the lower layers to desorb. When lighter volatile species are mixed with water below a crust, water reorganization  plays a key role in the sublimation process that could lead to a larger fraction of volatile compounds released when a thicker crust layer is present.  
\end{enumerate}

In this study, we finally investigated the desorption pattern of some volatile compounds in comet Hale Bopp \citep{biver20021995}. The slopes of the desorption of the volatile compounds have been attributed to sublimation dictated by water reorganization. This indicates that the volatile compounds trapped in the ice are heated to 140~K for the first time. The volatile compounds desorbing from the comet have been deposited at very low temperatures and are desorbing from amorphous water ice heated for the first time to high temperatures.

\begin{acknowledgements}
Funds from the Spanish MINECO PID2020-118974GB-C22 and PID2020-113084GB-I00  projects are acknowledged.
\end{acknowledgements}

%

\begin{thebibliography}{50}
\expandafter\ifx\csname natexlab\endcsname\relax\def\natexlab#1{#1}\fi

\bibitem[{{Altwegg} {et~al.}(2022){Altwegg}, {Combi}, {Fuselier}, {H{\"a}nni},
  {De Keyser}, {Mahjoub}, {M{\"u}ller}, {Pestoni}, {Rubin}, \&
  {Wampfler}}]{altwegg2022}
{Altwegg}, K., {Combi}, M., {Fuselier}, S.~A., {et~al.} 2022, \mnras, 516, 3900

\bibitem[{Ayotte {et~al.}(2001)Ayotte, Smith, Stevenson, Dohn{\'a}lek, Kimmel,
  \& Kay}]{ayotte2001effect}
Ayotte, P., Smith, R.~S., Stevenson, K.~P., {et~al.} 2001, Journal of
  Geophysical Research: Planets, 106, 33387

\bibitem[{Bergren {et~al.}(1978)Bergren, Schuh, Sceats, \&
  Rice}]{bergren1978oh}
Bergren, M.~S., Schuh, D., Sceats, M.~G., \& Rice, S.~A. 1978, The Journal of
  Chemical Physics, 69, 3477

\bibitem[{Biver {et~al.}(1997{\natexlab{a}})Biver, Bockel{\'e}e-Morvan, Colom,
  Crovisier, Davies, Dent, Despois, Gerard, Lellouch, Rauer,
  {et~al.}}]{biver1997evolution}
Biver, N., Bockel{\'e}e-Morvan, D., Colom, P., {et~al.} 1997{\natexlab{a}},
  Science, 275, 1915

\bibitem[{Biver {et~al.}(1997{\natexlab{b}})Biver, Bockel{\'e}e-Morvan, Colom,
  Crovisier, Germain, Lellouch, Davies, Dent, Moreno, Paubert,
  {et~al.}}]{biver1997long}
Biver, N., Bockel{\'e}e-Morvan, D., Colom, P., {et~al.} 1997{\natexlab{b}},
  Earth, Moon, and Planets, 78, 5

\bibitem[{Biver {et~al.}(2002)Biver, Bockel{\'e}e-Morvan, Colom, Crovisier,
  Henry, Lellouch, Winnberg, Johansson, Gunnarsson, Rickman,
  {et~al.}}]{biver20021995}
Biver, N., Bockel{\'e}e-Morvan, D., Colom, P., {et~al.} 2002, in Cometary
  Science after Hale-Bopp (Springer), 5--14

\bibitem[{Bockel{\'e}e-Morvan \& Biver(2017)}]{bockelee2017composition}
Bockel{\'e}e-Morvan, D. \& Biver, N. 2017, Philosophical Transactions of the
  Royal Society A: Mathematical, Physical and Engineering Sciences, 375,
  20160252

\bibitem[{Bockel{\'e}e-Morvan {et~al.}(2004)Bockel{\'e}e-Morvan, Crovisier,
  Mumma, \& Weaver}]{bockelee2004composition}
Bockel{\'e}e-Morvan, D., Crovisier, J., Mumma, M.~J., \& Weaver, H.~A. 2004,
  Comets II, 1, 391

\bibitem[{Bossa {et~al.}(2015)Bossa, Mat{\'e}, Fransen, Cazaux, Pilling, Rocha,
  Ortigoso, \& Linnartz}]{bossa2015porosity}
Bossa, J.-B., Mat{\'e}, B., Fransen, C., {et~al.} 2015, The Astrophysical
  Journal, 814, 47

\bibitem[{Brown \& Bolina(2007)}]{brown2007fundamental}
Brown, W.~A. \& Bolina, A.~S. 2007, Monthly Notices of the Royal Astronomical
  Society, 374, 1006

\bibitem[{Cazaux {et~al.}(2015)Cazaux, Bossa, Linnartz, \&
  Tielens}]{cazaux2015pore}
Cazaux, S., Bossa, J.-B., Linnartz, H., \& Tielens, A. G. G.~M. 2015, Astronomy
  \& Astrophysics, 573, A16

\bibitem[{Collings {et~al.}(2004)Collings, Anderson, Chen, Dever, Viti,
  Williams, \& McCoustra}]{collings2004laboratory}
Collings, M.~P., Anderson, M.~A., Chen, R., {et~al.} 2004, Monthly Notices of
  the Royal Astronomical Society, 354, 1133

\bibitem[{{Dartois} {et~al.}(2013){Dartois}, {Ding}, {de Barros}, {Boduch},
  {Brunetto}, {Chabot}, {Domaracka}, {Godard}, {Lv}, {Mej{\'{\i}}a Guam{\'a}n},
  {Pino}, {Rothard}, {da Silveira}, \& {Thomas}}]{dartois2013}
{Dartois}, E., {Ding}, J.~J., {de Barros}, A.~L.~F., {et~al.} 2013, Astronomy
  \& Astrophysics, 557, A97

\bibitem[{Dohn\'alek {et~al.}(2003)Dohn\'alek, Kimmel, Ayotte, Smith, \&
  Kay}]{dohnalek2003deposition}
Dohn\'alek, Z., Kimmel, G.~A., Ayotte, P., Smith, R.~S., \& Kay, B.~D. 2003,
  The Journal of Chemical Physics, 118, 364

\bibitem[{Dowell \& Rinfret(1960)}]{dowell1960low}
Dowell, L.~G. \& Rinfret, A.~P. 1960, Nature, 188, 1144

\bibitem[{Fulle {et~al.}(2010)Fulle, Colangeli, Agarwal, Aronica, Della~Corte,
  Esposito, Gr{\"u}n, Ishiguro, Ligustri, Moreno, {et~al.}}]{fulle2010comet}
Fulle, M., Colangeli, L., Agarwal, J., {et~al.} 2010, Astronomy \&
  Astrophysics, 522, A63

\bibitem[{Gundlach {et~al.}(2011)Gundlach, Skorov, \&
  Blum}]{gundlach2011outgassing}
Gundlach, B., Skorov, Y.~V., \& Blum, J. 2011, Icarus, 213, 710

\bibitem[{Hagen {et~al.}(1981)Hagen, Tielens, \& Greenberg}]{hagen1981infrared}
Hagen, W., Tielens, A. G. G.~M., \& Greenberg, J.~M. 1981, Chemical Physics,
  56, 367

\bibitem[{Hansen {et~al.}(2016)Hansen, Altwegg, Berthelier, Bieler, Biver,
  Bockel{\'e}e-Morvan, Calmonte, Capaccioni, Combi, De~Keyser,
  {et~al.}}]{hansen2016evolution}
Hansen, K.~C., Altwegg, K., Berthelier, J.-J., {et~al.} 2016, Monthly Notices
  of the Royal Astronomical Society, 462, S491

\bibitem[{He {et~al.}(2016)He, Acharyya, \& Vidali}]{he2016binding}
He, J., Acharyya, K., \& Vidali, G. 2016, The Astrophysical Journal, 825, 89

\bibitem[{He {et~al.}(2018)He, Emtiaz, \& Vidali}]{he2018measurements}
He, J., Emtiaz, S., \& Vidali, G. 2018, The Astrophysical Journal, 863, 156

\bibitem[{Isokoski {et~al.}(2014)Isokoski, Bossa, Triemstra, \&
  Linnartz}]{isokoski2014porosity}
Isokoski, K., Bossa, J.-B., Triemstra, T., \& Linnartz, H. 2014, Physical
  Chemistry Chemical Physics, 16, 3456

\bibitem[{Jenniskens {et~al.}(1997)Jenniskens, Banham, Blake, \&
  McCoustra}]{jenniskens1997liquid}
Jenniskens, P., Banham, S.~F., Blake, D., \& McCoustra, M.~R. 1997, The Journal
  of Chemical Physics, 107, 1232

\bibitem[{Jenniskens \& Blake(1994)}]{jenniskens1994structural}
Jenniskens, P. \& Blake, D.~F. 1994, Science, 265, 753

\bibitem[{Kossacki(2021)}]{kossacki2021sublimation}
Kossacki, K.~J. 2021, Icarus, 368, 114613

\bibitem[{Kossacki {et~al.}(2017)Kossacki, Leliwa-Kopystynski, Witek, Jasiak,
  \& Dubiel}]{kossacki2017sublimation}
Kossacki, K.~J., Leliwa-Kopystynski, J., Witek, P., Jasiak, A., \& Dubiel, A.
  2017, Icarus, 294, 227

\bibitem[{Krause {et~al.}(2011)Krause, Blum, Skorov, \&
  Trieloff}]{krause2011thermal}
Krause, M., Blum, J., Skorov, Y.~V., \& Trieloff, M. 2011, Icarus, 214, 286

\bibitem[{Laufer {et~al.}(1987)Laufer, Kochavi, \&
  Bar-Nun}]{laufer1987structure}
Laufer, D., Kochavi, E., \& Bar-Nun, A. 1987, Physical Review B, 36, 9219

\bibitem[{Linstrom \& Mallard(2001)}]{linstrom2001nist}
Linstrom, P.~J. \& Mallard, W.~G. 2001, Journal of Chemical \& Engineering
  Data, 46, 1059

\bibitem[{Luna {et~al.}(2014)Luna, Satorre, Santonja, \& Domingo}]{luna2014new}
Luna, R., Satorre, M.~{\'A}., Santonja, C., \& Domingo, M. 2014, Astronomy \&
  Astrophysics, 566, A27

\bibitem[{Marboeuf {et~al.}(2012)Marboeuf, Schmitt, Petit, Mousis, \&
  Fray}]{marboeuf2012cometary}
Marboeuf, U., Schmitt, B., Petit, J.-M., Mousis, O., \& Fray, N. 2012,
  Astronomy \& Astrophysics, 542, A82

\bibitem[{Mart{\'\i}n-Dom{\'e}nech {et~al.}(2014)Mart{\'\i}n-Dom{\'e}nech,
  Mu{\~n}oz~Caro, Bueno, \& Goesmann}]{martin2014thermal}
Mart{\'\i}n-Dom{\'e}nech, R., Mu{\~n}oz~Caro, G.~M., Bueno, J., \& Goesmann, F.
  2014, Astronomy \& Astrophysics, 564, A8

\bibitem[{Mastrapa {et~al.}(2009)Mastrapa, Sandford, Roush, Cruikshank, \&
  Dalle~Ore}]{mastrapa2009optical}
Mastrapa, R.~M., Sandford, S.~A., Roush, T.~L., Cruikshank, D.~P., \&
  Dalle~Ore, C.~M. 2009, The Astrophysical Journal, 701, 1347

\bibitem[{Mat{\'e} {et~al.}(2021)Mat{\'e}, Carrasco-Herrera, Tim{\'o}n,
  Tanarro, Herrero, Carrascosa, Caro, Gonz{\'a}lez-D{\'\i}az, \&
  Jim{\'e}nez-Serra}]{mate20212}
Mat{\'e}, B., Carrasco-Herrera, R., Tim{\'o}n, V., {et~al.} 2021, The
  Astrophysical Journal, 909, 123

\bibitem[{Mat{\'e} {et~al.}(2020)Mat{\'e}, Cazaux, Satorre, Molpeceres,
  Ortigoso, Mill{\'a}n, \& Santonja}]{mate2020diffusion}
Mat{\'e}, B., Cazaux, S., Satorre, M.~{\'A}., {et~al.} 2020, Astronomy \&
  Astrophysics, 643, A163

\bibitem[{Maté {et~al.}(2023)Maté, Tanarro, \& Timón}]{submittedBelen}
Maté, B., Tanarro, I., \& Timón, V. 2023, Monthly Notices of the Royal
  Astronomical Society, submitted

\bibitem[{May {et~al.}(2013a)May, Smith, \& Kay}]{may2013release}
May, R.~A., Smith, R.~S., \& Kay, D.~B. 2013a, The Journal of Chemical Physics,
  138, 104501

\bibitem[{May {et~al.}(2013b)May, Smith, \& Kay}]{alan2013release}
May, R.~A., Smith, R.~S., \& Kay, D.~B. 2013b, The Journal of Chemical Physics,
  138, 104502

\bibitem[{Meech \& Svoren(2004)}]{meech2004using}
Meech, K. \& Svoren, J. 2004, Comets II, 1, 317

\bibitem[{Minissale {et~al.}(2022)Minissale, Aikawa, Bergin, Bertin, Brown,
  Cazaux, Charnley, Coutens, Cuppen, Guzman, {et~al.}}]{minissale2022thermal}
Minissale, M., Aikawa, Y., Bergin, E., {et~al.} 2022, ACS Earth and Space
  Chemistry, 6, 597

\bibitem[{Molpeceres {et~al.}(2017)Molpeceres, Satorre, Ortigoso, Zanchet,
  Luna, Mill{\'a}n, Escribano, Tanarro, Herrero, \&
  Mat{\'e}}]{molpeceres2017physical}
Molpeceres, G., Satorre, M.~{\'A}., Ortigoso, J., {et~al.} 2017, Monthly
  Notices of the Royal Astronomical Society, 466, 1894

\bibitem[{{Mumma} {et~al.}(2002){Mumma}, {Disanti}, {dello Russo},
  {Magee-Sauer}, {Gibb}, \& {Novak}}]{2002ESASP.500..753M}
{Mumma}, M.~J., {Disanti}, M.~A., {dello Russo}, N., {et~al.} 2002, in ESA
  Special Publication, Vol. 500, Asteroids, Comets, and Meteors: ACM 2002, ed.
  B.~{Warmbein}, 753--762

\bibitem[{Mumma {et~al.}(1996)Mumma, DiSanti, Russo, Fomenkova, Magee-Sauer,
  Kaminski, \& Xie}]{mumma1996detection}
Mumma, M.~J., DiSanti, M.~A., Russo, N.~D., {et~al.} 1996, Science, 272, 1310

\bibitem[{Pat-El {et~al.}(2009)Pat-El, Laufer, Notesco, \&
  Bar-Nun}]{pat2009experimental}
Pat-El, I., Laufer, D., Notesco, G., \& Bar-Nun, A. 2009, Icarus, 201, 406

\bibitem[{{Prialnik} {et~al.}(2004){Prialnik}, {Benkhoff}, \&
  {Podolak}}]{prialnick2004}
{Prialnik}, D., {Benkhoff}, J., \& {Podolak}, M. 2004, Comets II, 1, 359

\bibitem[{Raut {et~al.}(2007)Raut, Teolis, Loeffler, Vidal, Fam{\'a}, \&
  Baragiola}]{raut2007compaction}
Raut, U., Teolis, B., Loeffler, M., {et~al.} 2007, The Journal of Chemical
  Physics, 126, 244511

\bibitem[{Rubin {et~al.}(2020)Rubin, Engrand, Snodgrass, Weissman, Altwegg,
  Busemann, Morbidelli, \& Mumma}]{rubin2020origin}
Rubin, M., Engrand, C., Snodgrass, C., {et~al.} 2020, Space science reviews,
  216, 1

\bibitem[{Smith {et~al.}(2016)Smith, May, \& Kay}]{smith2016desorption}
Smith, R.~S., May, R.~A., \& Kay, B.~D. 2016, The Journal of Physical Chemistry
  B, 120, 1979

\bibitem[{Snodgrass {et~al.}(2016)Snodgrass, Opitom, de~Val-Borro, Jehin,
  Manfroid, Lister, Marchant, Jones, Fitzsimmons, Steele,
  {et~al.}}]{snodgrass2016perihelion}
Snodgrass, C., Opitom, C., de~Val-Borro, M., {et~al.} 2016, Monthly Notices of
  the Royal Astronomical Society, 462, S138

\bibitem[{Stern {et~al.}(1999)Stern, Colwell, Festou, Tamblyn, Parker, Slater,
  Weissman, \& Paxton}]{stern1999comet}
Stern, S.~A., Colwell, W.~B., Festou, M.~C., {et~al.} 1999, The Astronomical
  Journal, 118, 1120

\end{thebibliography}
%

\end{document}